\documentclass[12pt]{article}
\usepackage{graphicx}

\usepackage[colorlinks=true, urlcolor=blue, citecolor=black]{hyperref}  

\newcommand\pubnumber{DPF2015-58}
\newcommand\pubdate{\today}

\def\osu{Department of Physics \\
The Ohio State University, Columbus, Ohio 43210}
\def\support{\footnote{On behalf of the CMS Collaboration}}

\def\Title#1{\begin{center} {\Large #1 } \end{center}}
\def\Author#1{\begin{center}{ \sc #1} \end{center}}
\def\Address#1{\begin{center}{ \it #1} \end{center}}

\newcommand\pubblock{\rightline{\begin{tabular}{l} \pubnumber\\
         \pubdate  \end{tabular}}}
\newenvironment{Abstract}{\begin{quotation}  }{\end{quotation}}
\newenvironment{Presented}{\begin{quotation} \begin{center} 
             PRESENTED AT\end{center}\bigskip 
      \begin{center}\begin{large}}{\end{large}\end{center} \end{quotation}}
\def\Acknowledgments{\bigskip  \bigskip \begin{center} \begin{large}
             \bf ACKNOWLEDGMENTS \end{large}\end{center}}

\def\beq{\begin{equation}}
\def\eeq#1{\label{#1}\end{equation}}
\def\eeqn{\end{equation}}

\def\beqa{\begin{eqnarray}}
\def\eeqa#1{\label{#1}\end{eqnarray}}
\def\eeqan{\end{eqnarray}}

\let\bar=\overbar

\def\Dslash{\not{\hbox{\kern-4pt $D$}}}
\def\dslash{\not{\hbox{\kern-2pt $\del$}}}

\def\msb{{\bar{\ssstyle M \kern -1pt S}}}

\begin{document}
\begin{titlepage}
\pubblock

\vfill
\Title{Performance of the Cathode Strip Chamber endcap muon detectors in Run 2}
\vfill
\Author{ H. Wells Wulsin \support}
\Address{\osu}
\vfill
\begin{Abstract}
Since the end of Run 1 of the LHC in 2012, the outermost ring has been added to the CMS Cathode Strip Chambers (CSC) endcap muon detector, and the readout electronics of the innermost ring of CSCs have been upgraded to accommodate the larger luminosity and collision energy anticipated in Run 2. A major effort was required to build, install, and commission these new chambers and electronics. This talk summarizes the improvements made during this upgrade and presents the performance of the CSC detector during the early stages of Run 2.  
\end{Abstract}
\vfill
\begin{Presented}
DPF 2015\\
The Meeting of the American Physical Society\\
Division of Particles and Fields\\
Ann Arbor, Michigan, August 4--8, 2015\\
\end{Presented}
\vfill
\end{titlepage}
\def\thefootnote{\fnsymbol{footnote}}
\setcounter{footnote}{0}

\section{Introduction}

Muons are critical to the physics program at CMS.  Three different technologies are employed in the CMS muon detectors:  drift tubes located in the barrel region, resistive plate chambers in the barrel and endcaps, and cathode strip chambers (CSCs) located in the endcaps.  The CSCs, the focus of this talk, 
record muons with pseudorapidity $0.9 < |\eta| < 2.4$.  A detector in this forward region confronts challenges including 
a non-uniform magnetic field and large radiation doses.  

In Run 1 the CSC system consisted of 468 multi-wire proportional chambers, each of which was constructed from seven copperclad trapezoidal panels, forming 6 gas-gap layers.  Cathode strips, 3-16 mm wide, are etched onto the copper panels and measure the azimuthal coordinate of a muon, which is critical for the measurement of its transverse momentum.  Between each layer of cathode strips, anode wires are aligned perpendicular to the strips.  The wires are held at a potential of 2.9 - 3.6 kV, and provide the radial coordinate of the muon as well as a precise measurement of its timing.  

Three-dimensional segments are obtained by fitting the hits measured in each of the six layers.  These segments, known as local charged tracks, are used as primitive objects for triggering events of interest.  

The CSCs in each endcap are grouped into four stations with varying longitudinal distance to the proton-proton interaction point.   Within each station, there are two or three rings of chambers, with varying radial distance from the beamline.  The chambers are labelled MEX/Y/Z, for Muon Endcap chambers in station X, ring Y, and azimuthal position Z.  

\section{Upgrades in Long Shutdown 1}

There were two main CSC upgrade projects during the LHC Long Shutdown 1, from 2013-2014.  In the outermost ring, ME4/2, 72 new chambers, which were missing in Run 1, were built, installed, and commissioned.  In the innermost ring, ME1/1, which faces the largest flux of muons of all the CSCs, the readout electronics were replaced to cope with large rates.

The installation of the new ME4/2 chambers provides four-station redundancy for muons with $|\eta| < 2.4$.  In particular, it improves the efficiency of muon reconstruction in the region of $1.2 < |\eta| < 1.8$ 
and reduces the fake rate for triggering and for offline reconstruction.  
This permits reduced momentum thresholds for muon triggers.  
The panels for these chambers were cut, milled, cleaned, and polished at Fermilab.  The chambers were 
constructed at CERN, which included the wire winding, assembly, and instrumentation.  

The ME1/1 chambers are sensitive to muons in the range of $1.6 < \eta < 2.4$.  They are segmented into two sub-chambers, ME1/1a and ME1/1b, by a gap along the cathode strips.  The wires have different spacings and orientation in the two regions.  In Run 1, the ME1/1a strips were triple-ganged every 16 strips, which caused a three-fold ambiguity in triggering and reconstruction.  This was chosen to reduce the number of readout channels.  
During this upgrade, two additional on-chamber readout boards have been added so that each strip will be read out 
by a single channel.  The ME1/1 upgrade involved replacing electronics boards on the chambers, in peripheral crates attached to the CMS iron disk, and off the detector.  A detailed description of the electronics used in Run 1 is given 
in~\cite{Bylsma}.  

On the chambers, the five Cathode Front End Boards (CFEBs) used in Run 1 were replaced with seven Digital CFEBs.  
Each DCFEB has over 2400 individual electronics components.  They have a fully digital pipeline, which has effectively no deadtime, and thus is better able to cope with events of high pile-up.  The DCFEBs also feature a newer FPGA, the Virtex 6, and optical data read out.  For this upgrade, 554 DCFEBs, including 50 spare boards, were produced and tested.  The old CFEB's were installed onto the new ME4/2 chambers.  The low voltage distribution and monitoring boards on the chambers were also improved to handle the increased power consumption and number of channels of the seven DCFEBs.  Also, the anode local charge track board on the chamber was equipped with a new FPGA, the Spartan 6, to handle the higher rates.  

In the peripheral crates, the Data acquisition Mother Board (DMB) and the Trigger Mother Board (TMB) were replaced to accept optical inputs from the DCFEBs.  The DMB has two main tasks, distribution of trigger, timing and control signals to the chambers, and acquisition of data from the chambers.  The upgraded Optical DMB has an improved FPGA, the Virtex 6, with new firmware to handle seven DCFEBs.  The TMB constructs cathode track segments and performs matching to anode tracks for triggering.  The new Optical TMB features a mezzanine board with a Virtex 6 FPGA and improved trigger logic.  A patch panel interface board has been added to enable communication between the Optical DMB and the seven DCFEBs.  

Off the detector, behind the radiation wall of the experimental cavern, the Detector Dependent Unit boards were modified to send send data directly to the central CMS data acquisition via optical link.  

The ME4/2 and ME1/1 chambers were subjected to a comprehensive battery of tests on the surface before being installed on CMS in the underground
experimental cavern.  These included tests of basic functionality, connectivity, trigger logic, data quality, and noise levels.   Some tests use generated pulses while others test reconstruction of cosmic ray muons.  Each chamber was also 
subjected to four weeks of high-voltage monitoring and gas leak tests.  After installation on the detector, tests were 
repeated in situ.  Finally, the CSCs were included with the other CMS subdetectors for collection of cosmic ray muon data.  During the course of this testing, a few problems with the chambers were identified and repaired, such as faulty electronics boards or loose cable connections.  

\section{CSC Performance in Run 2}
The first stable proton-proton collisions at $\sqrt{s} = 13$ TeV were delivered by the LHC on 3 June 2015.  Since that time the CSCs have performed well, exceeding the level achieved at the end of Run 1.  

\begin{figure}[!h]
\centering
\begin{minipage}{0.45\linewidth}
\centerline{\hskip1.5cm Run 1}
\centerline{\includegraphics[width=\linewidth]{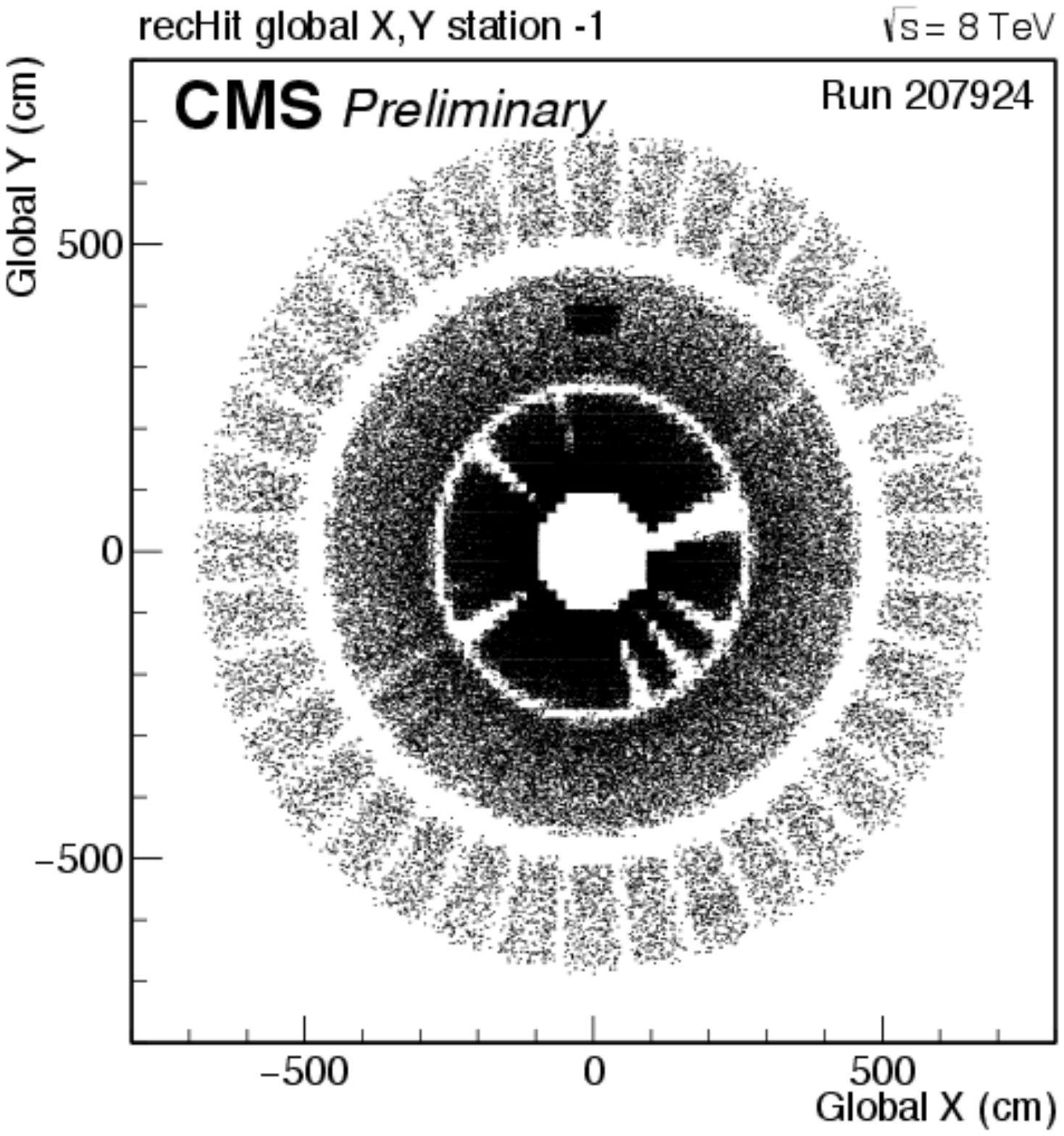}}
\centerline{\includegraphics[width=\linewidth]{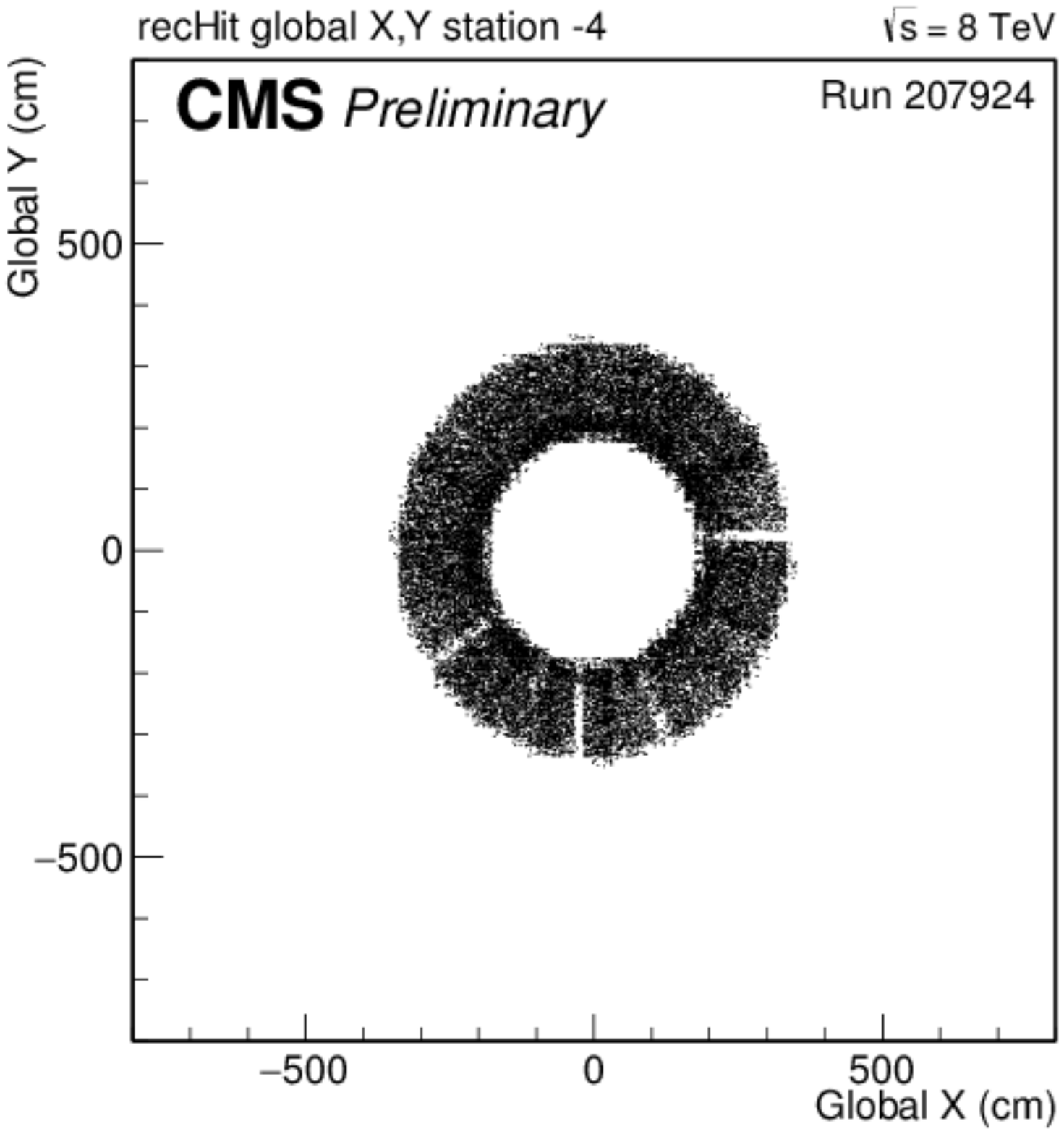}}
\end{minipage}
\hfill
\begin{minipage}{0.45\linewidth}
\centerline{\hskip1.5cm Run 2}
\centerline{\includegraphics[width=\linewidth]{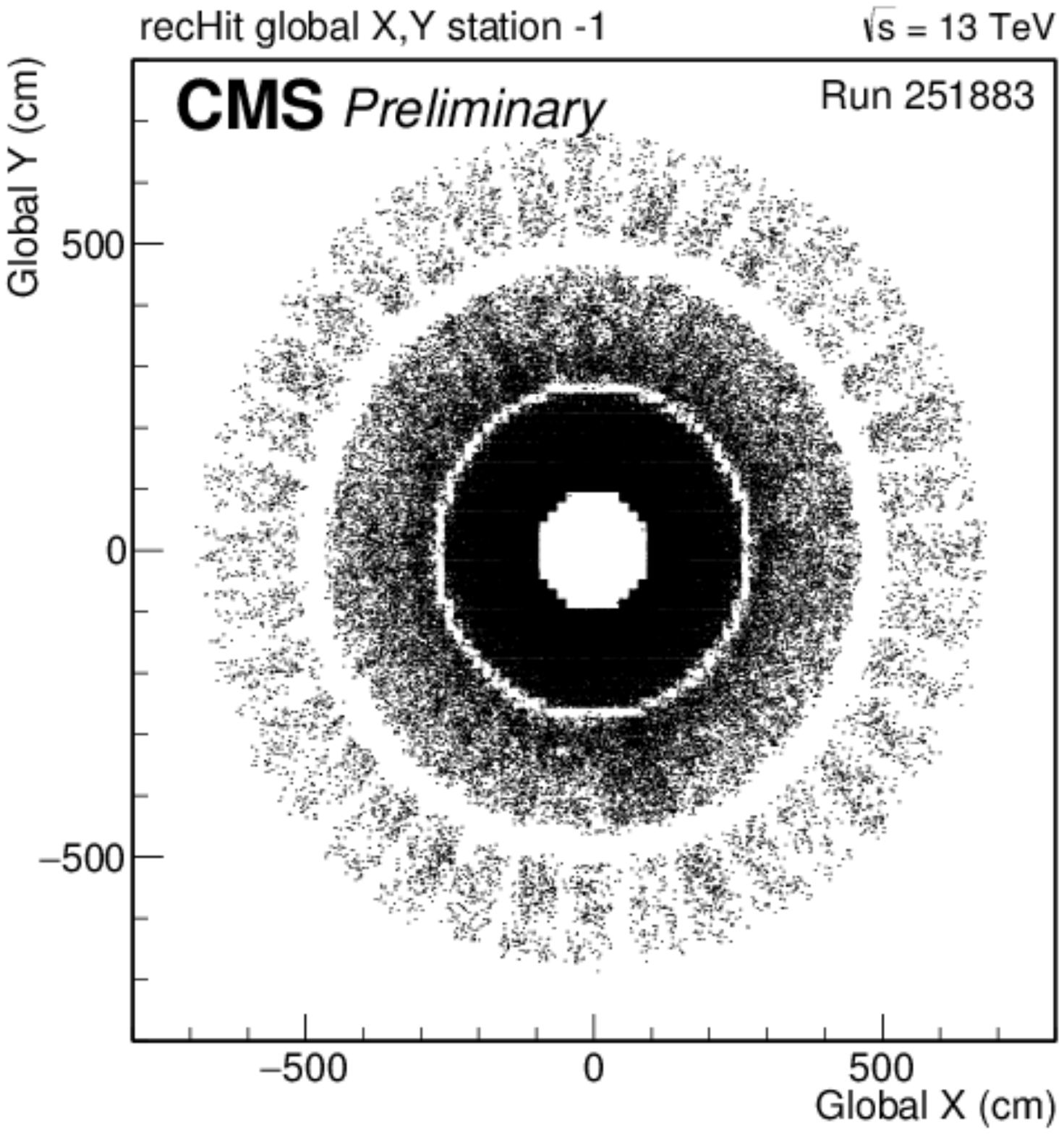}}
\centerline{\includegraphics[width=\linewidth]{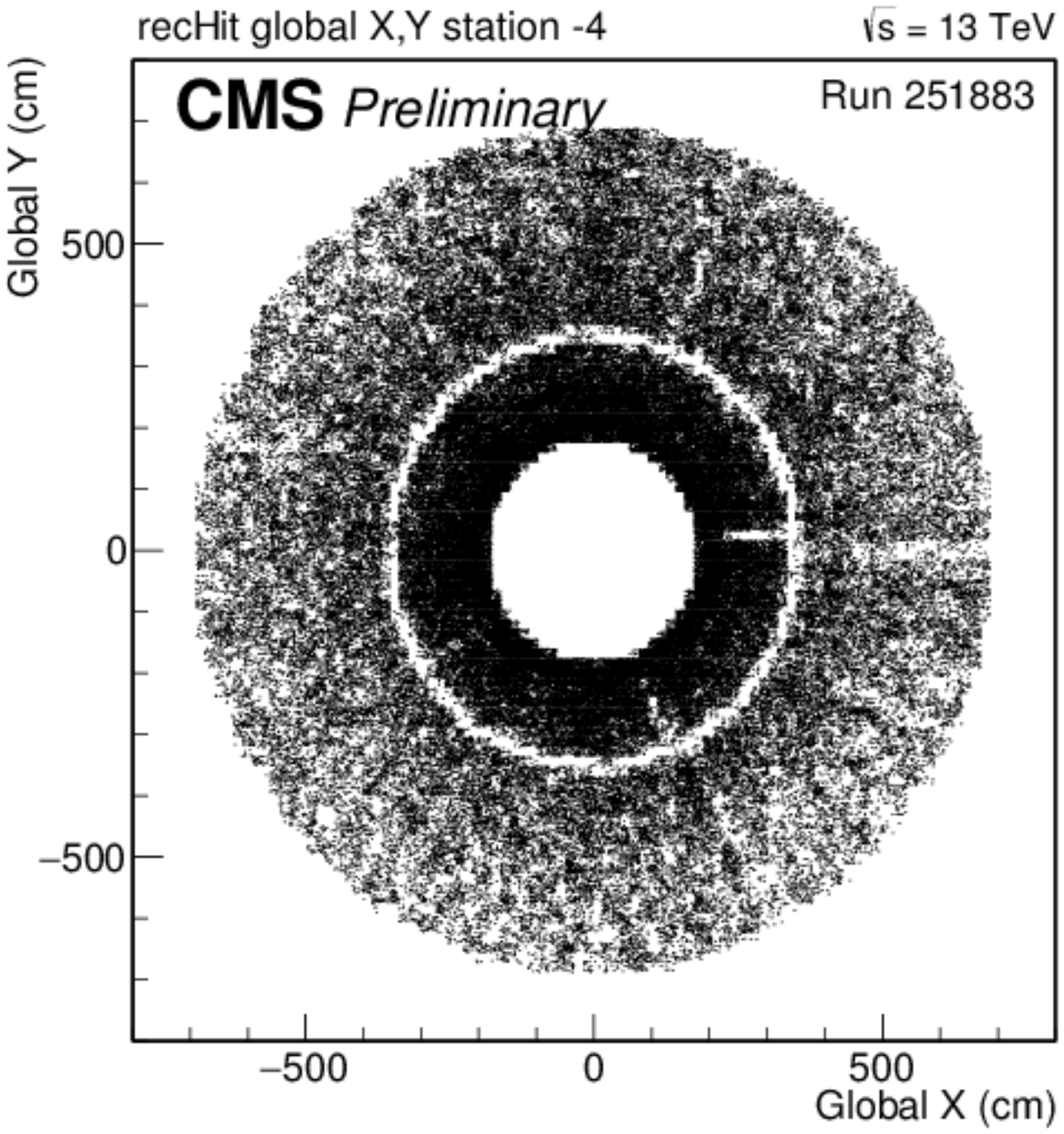}}
\end{minipage}
\caption{The hit reconstruction efficiency has improved from Run 1 (left) to Run 2 (right), as a result of the repair of several problematic ME1/1 chambers (top) and the addition of the ME4/2 chambers (bottom).  }  
\label{fig:hitEff}
\end{figure}

The occupancy of reconstructed hits is uniform across chambers in a given ring.  This is an important test, since problematic chambers typically produce low or high occupancy.  The hit reconstruction efficiency has also improved, as shown in Figure~\ref{fig:hitEff}, thanks to the repair of several problematic ME1/1 chambers and the addition of the ME4/2 chambers.  

\begin{figure}[!h]
\centering
\centerline{\includegraphics[width=0.6\linewidth]{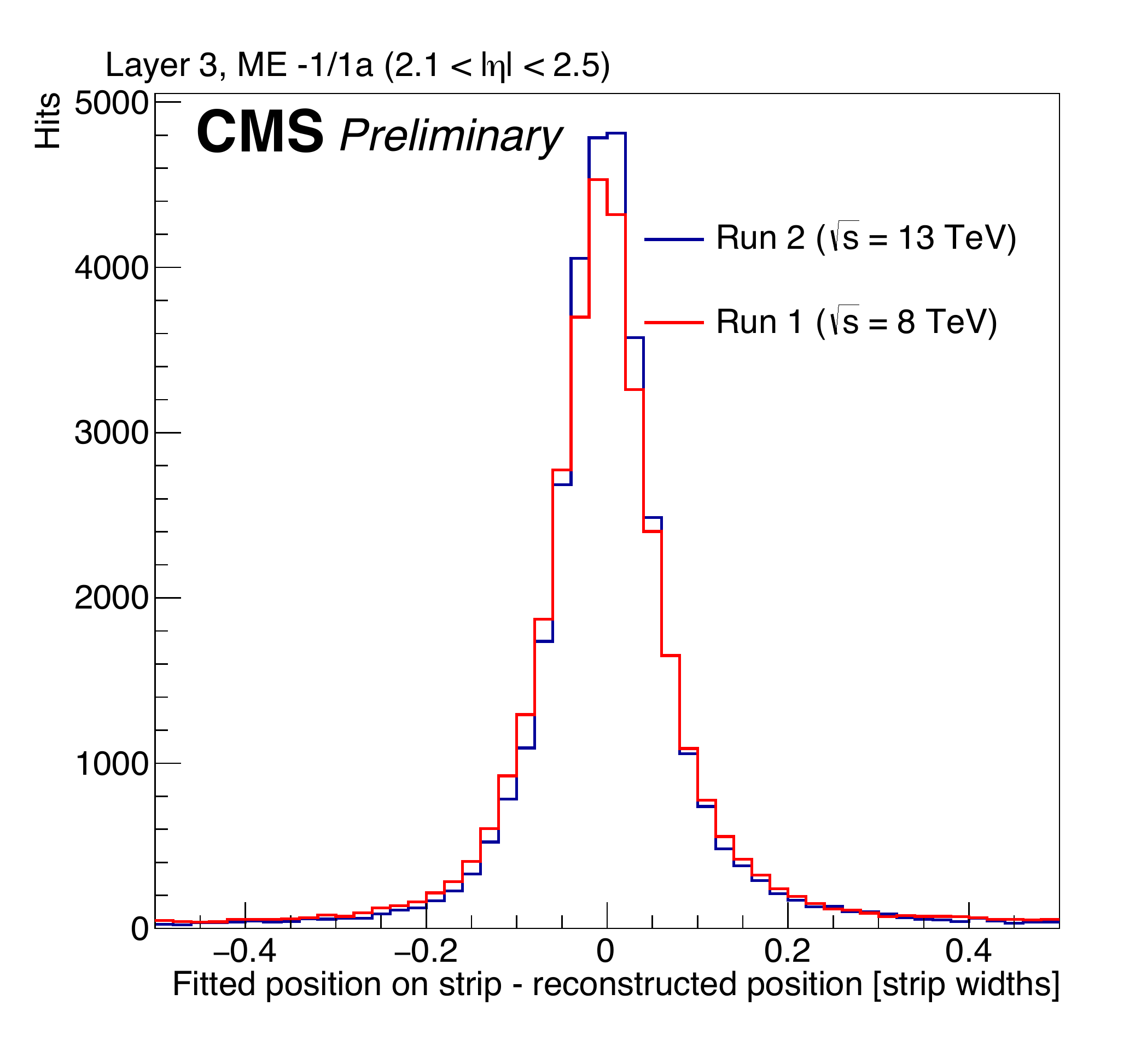}}
\caption{The hit position resolution has improved from 64 $\mu$m in Run 1 to 51 $\mu$m in Run 2.}  
\label{fig:hitRes}
\end{figure}

The position resolution of hits recorded in ME1/1a has improved by about 20\%, from 64 $\mu$m in Run 1 to 51 $\mu$m in Run 2, as illustrated in Figure~\ref{fig:hitRes}.  The resolution is measured by finding the difference between the position of a reconstructed hit in a single layer and the position obtained by fitting a segment with hits from the other five layers.  This improvement in resolution is primarily from the removal of triple-ganging the strips. This reduces the capacitance, which in turn decreases the front-end noise.  The resolution in all CSC stations varies from 50--140 $\mu$m, depending on the station, which corresponds to an azimuthal resolution of $\sim$0.1 mrad.  

\begin{figure}[!h]
\centering
\begin{minipage}{0.45\linewidth}
\centerline{\includegraphics[width=\linewidth]{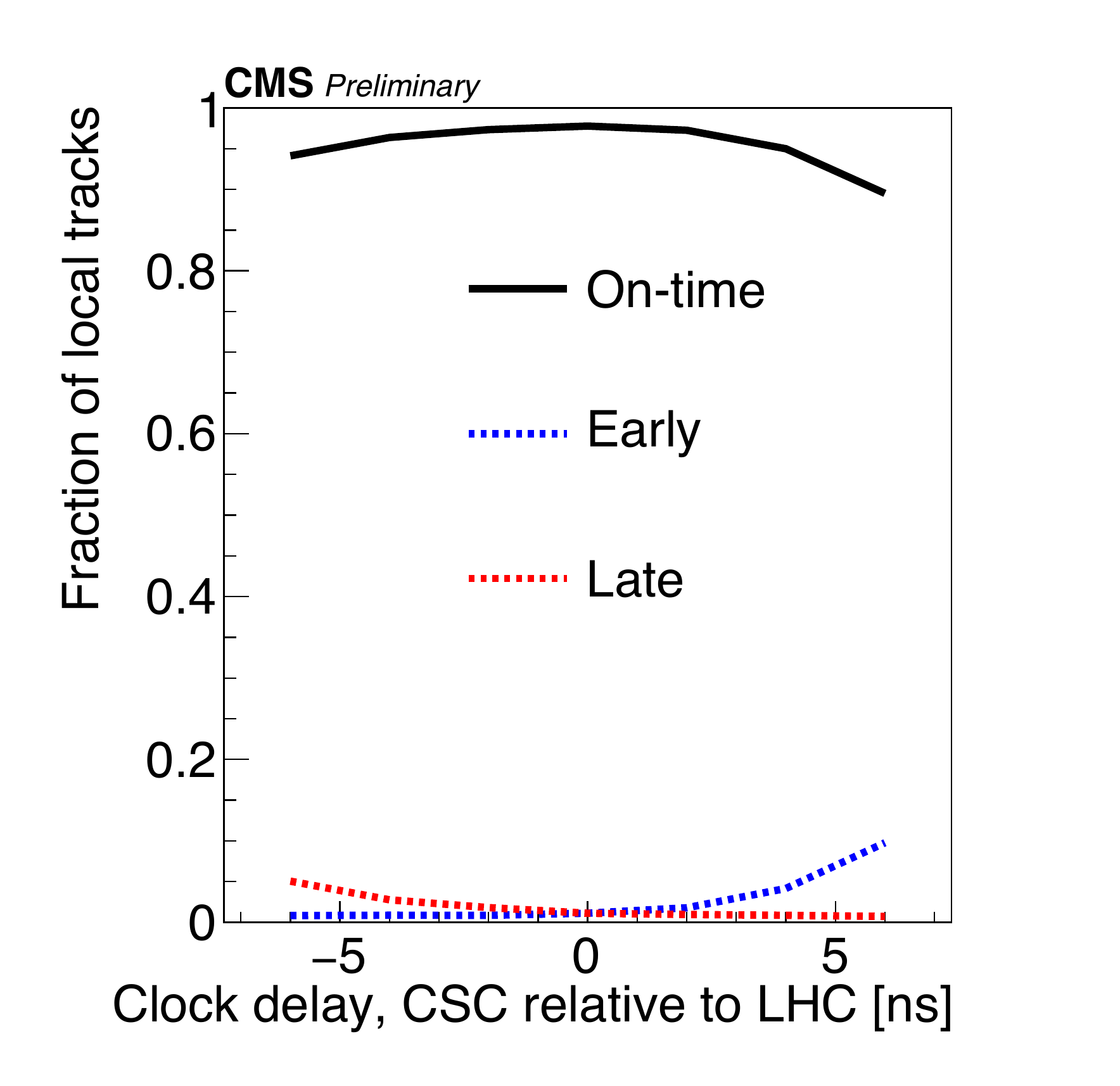}}
\end{minipage}
\hfill
\begin{minipage}{0.45\linewidth}
\centerline{\includegraphics[width=\linewidth]{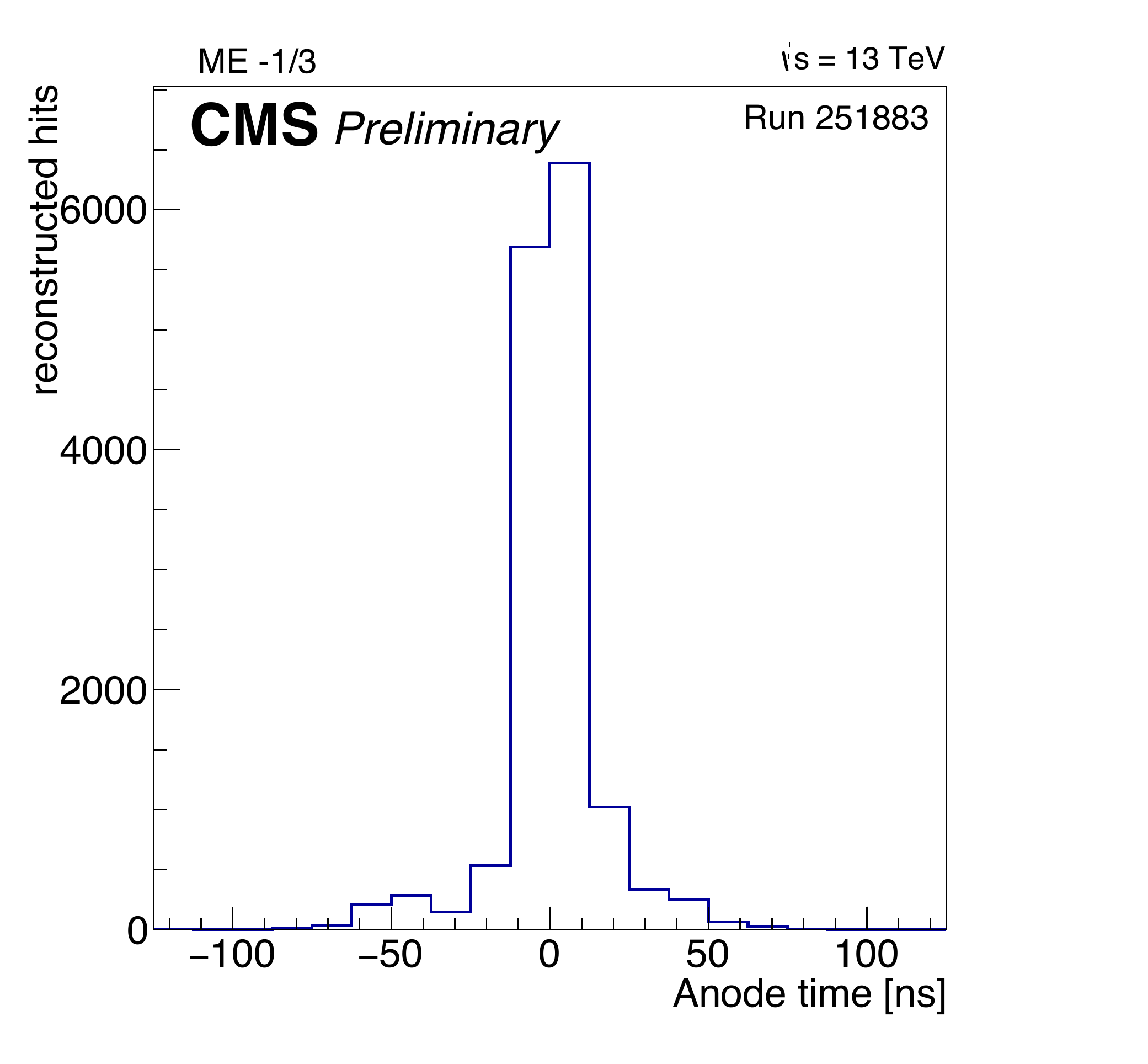}}
\end{minipage}
\caption{
The fraction of local tracks arriving on-time, early, and late for various clock delays (left).  The interaction point time measured by the anode wires for single reconstructed hits is centered at 0 (right), as expected for timed-in chambers. }
\label{fig:timing}
\end{figure}

The CSCs have been timed in by varying the CSC clock delay relative to the LHC clock, and choosing the delay that maximizes the number of local tracks in the correct time window.  The fraction of local tracks arriving on-time, early, and late is shown for various clock delays in Figure~\ref{fig:timing}.  
That figure also shows that the interaction point time measured by the anode wires for single reconstructed hits is centered at 0, as expected for timed-in chambers. The timing and synchronization accuracy is at the level of Run 1.

\section{CMS Muon Reconstruction in Run 2}
In Run 2, the CMS muon system as a whole has performed well.  The efficiency of muon reconstruction has improved by 1--2\% by implementing reconstruction algorithms that use new muon-specific tracking iterations.  This improvement is independent of pseudorapidity and the number of primary vertices in the event.  The simulation of muon reconstruction agrees well with the data, as measured in $Z\rightarrow\mu\mu$ events.  Using 20 pb$^{-1}$ of data taken at $\sqrt{s}=13$ TeV, various known dimuon resonances have been observed, including the Z, J/$\psi$, $\phi$ and $\Upsilon$.  

\section{Conclusion} 
During the LHC Long Shutdown 1, a great effort of many people went toward preparing the CSCs for Run 2.  Improved ME1/1 electronics remove the ambiguity from triple-ganging the strips in ME1/1a and reduce deadtime by storing data in a digital pipeline.  New ME4/2 chambers complete the fourth station, resulting in reduced fake rates and improved efficiency.  The CSCs have been timed in and are taking data with high efficiency.  The performance of the CSCs in the early stages is improved relative to the end of Run 1.

\Acknowledgments
I thank the many members of the CSC project who have contributed to the upgrade, maintenance, and operation of this detector.  I also thank all of our CMS colleagues, the CERN accelerator division for the excellent operation of the LHC, and the many funding agencies that support our work.


\begin{thebibliography}{99}


\bibitem{Bylsma} Bylsma, B.G., et al., "The cathode strip chamber data acquisition electronics for CMS", \href{https://inis.iaea.org/search/search.aspx?orig_q=RN:41029089}{{\em NIM} A 600, 661 (2009)}.  


\end{thebibliography}
\end{document}